\documentclass[conference]{IEEEtran}
\IEEEoverridecommandlockouts

\usepackage{cite}
\usepackage{amsmath,amssymb,amsfonts,bm}
\usepackage{makecell,arydshln,booktabs,url}
\usepackage{algorithmic}
\usepackage{graphicx}
\usepackage{textcomp}
\usepackage{xcolor}
\def\BibTeX{{\rm B\kern-.05em{\sc i\kern-.025em b}\kern-.08em
    T\kern-.1667em\lower.7ex\hbox{E}\kern-.125emX}}
\begin{document}

\title{Transcribing and Translating, Fast and Slow:\\ Joint Speech Translation and Recognition\\
\thanks{*equal contribution}
}

\author{\IEEEauthorblockN{Niko Moritz*, Ruiming Xie*, Yashesh Gaur, Ke Li, Simone Merello \\ Zeeshan Ahmed, Frank Seide, Christian Fuegen}
\IEEEauthorblockA{\textit{Meta AI}}}

\maketitle
\begin{abstract}

We propose the joint speech translation and recognition (JSTAR) model that leverages the fast-slow cascaded encoder architecture for simultaneous end-to-end automatic speech recognition (ASR) and speech translation (ST).
The model is transducer-based and uses a multi-objective training strategy that optimizes both ASR and ST objectives simultaneously. This allows JSTAR to produce high-quality streaming ASR and ST results. We apply JSTAR in a bilingual conversational speech setting with smart-glasses, where the model is also trained to distinguish speech from different directions corresponding to the wearer and a conversational partner.
Different model pre-training strategies are studied to further improve results, including training of a transducer-based streaming machine translation (MT) model for the first time and applying it for parameter initialization of JSTAR.
We demonstrate superior performances of JSTAR compared to a strong cascaded ST model in both BLEU scores and latency.

\end{abstract}
\section{Introduction}
\label{sec:intro}

Streaming ST is a challenging task that has received increasing attention in recent years \cite{seamless2023, papi2024streamatt, zhang2024streamspeech, parnia2019study}.
Conventional approaches often rely on cascaded systems, where ASR and MT are performed sequentially. While this modular approach allows for easy utilization of large text datasets for MT training, it requires complex beam search algorithms in streaming applications \cite{rabatin2024}. This can lead to increased latency and a loss of translation accuracy due to error propagation.
Alternatively, end-to-end speech-to-text translation (S2TT) systems can avoid the aforementioned problems.

Prior art in this field includes several approaches aimed at improving efficiency and accuracy \cite{liu19d_interspeech,xian2021_multilingual,NEURIPS2023_b6262f7a}.
We primarily focus on RNN-T based architectures due to their advantages in streaming applications \cite{Graves12rnnt, He2019e2eASRmobile, Moritz_DCN}. For example, \cite{papi2023sot} proposes a S2TT system that uses serialized output training (SOT) \cite{Chang2022extGTC,kanda2022streaming} with the RNN-T loss to produce the translation text immediately after the ASR output in the same transcription sequence.
Another approach proposed in \cite{liu2021} extends the joiner of an RNN-T model with cross-attention for S2TT, which is to allow better word reorderings for the translation task.
In \cite{Xue2022LargeScaleSE}, an extension of \cite{liu2021} is proposed where instead of augmenting the joiner with cross-attention, an attention pooling mechanism is introduced that can be seen as a gating mechanism to better fuse encoder and decoder information per frame for the translation task.
In \cite{wang2022lamassu}, the authors propose the Streaming Language-Agnostic Multilingual Speech Recognition and Translation Model Using Neural Transducers (LAMASSU) approach, which is a transducer-based model capable of both streaming ASR and speech translation (ST).
To generate multilingual output, two architecture alternatives are studied: using separate joiner and prediction networks for each target language or using a single unified joiner and prediction network with a target language ID token. Additionally, a clustered multi-lingual encoder is proposed, which includes transformer modules with shared and parallel layers that are trained to specialize for a particular language using an auxiliary loss function.

In this paper, we introduce JSTAR, a novel approach for simultaneous speech recognition and translation. JSTAR leverages an RNN-T based cascaded fast-slow encoder architecture \cite{Mahadeokar2022StreamingPT}, which in this work is equipped with two joiner and predictor networks.
JSTAR employs a multi-objective training strategy that optimizes both ASR and ST objectives simultaneously. The ASR task utilizes a joiner and predictor applied to the fast encoder, which is based on a low-latency ConvEmformer setup \cite{Yangyang2022convemformer}. In contrast, the ST task leverages the output of the slow encoder, which has twice the context size, and applies a separate joiner and predictor to it.
The fast-slow architecture enables JSTAR to effectively balance the trade-off between latency and accuracy of both tasks, since accurate ST typically requires a broader contextual understanding.
We apply JSTAR to recognize and transcribe bilingual conversations with smart glasses using a multi-channel directional ASR solution with speaker attribution capabilities \cite{Lin2024AGADIR, lin2023directional}.
To demonstrate the effectiveness of our approach, we conduct extensive experiments using FLEURS \cite{conneau2023fleurs} with multi-channel and multi-talker data simulation, as well as using realistic recordings of bilingual speech conversations.
In addition, for the first time we propose to train a streaming MT model based on RNN-T, which we leverage for model parameter initialization of JSTAR.

\section{Joint Speech Translation and Recognition}
\label{sec:jstar}

In this section, the JSTAR model for simultaneous ASR and ST is described for an input sequence $X = (\bm{x}_1, \dots, \bm{x}_T)$ of length $T$,
where $\bm{x}_t = (x_1, \dots, x_C)$ denotes a vector of a multi-channel audio sample at time index $t$ received from an array of $C$ microphones with $x_c$ being a single audio sample for channel index $c$.
Label sequence $Y = (y_1, \dots, y_U)$ denotes an ASR label sequence of length $U$ with label index $u$ and $Z = (z_1, \dots, z_L)$ denotes the ST label sequence of length $L$ for the translated text using $u$ as the label index.
Both $Y$ and $Z$ correspond to a sequence of words or word-pieces.

\subsection{Model Architecture}
\label{ssec:model_arch}

\begin{figure}[tb]
  \centering
  \centerline{\includegraphics[width=0.7\linewidth]{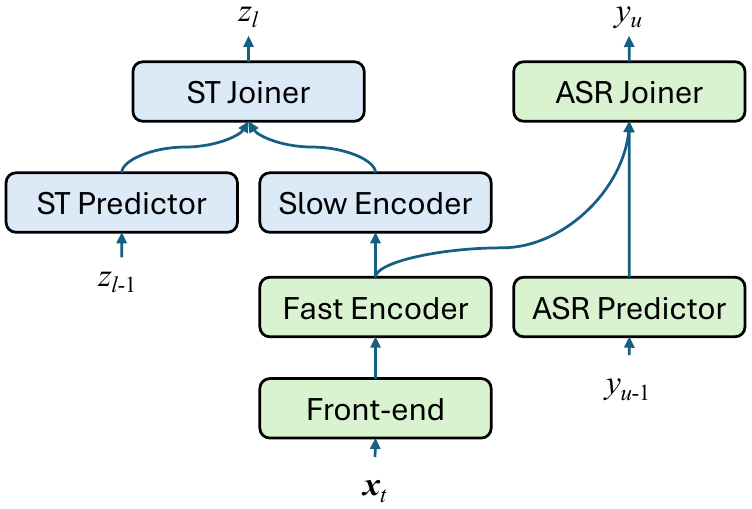}}
  \vspace{-2mm}
  \caption{JSTAR model architecture.}
  \vspace{-5mm}
\label{fig:jstar_arch}
\end{figure}

The proposed model architecture is RNN-T based and uses a cascaded fast and slow encoder architecture \cite{Mahadeokar2022StreamingPT}.
In contrast to the original fast and slow architecture, however, separate predictor and joiner modules are used for the ASR and ST tasks as shown in Figure~\ref{fig:jstar_arch}.
The RNN-T objective is to minimize the loss function
\begin{equation}
\label{eq:rnnt}
    \vspace{-0.8mm}
    \mathcal{L}_\text{asr} = - \ln \sum_{\pi \in \mathcal{B}^{-1}(Y)} p(\pi | X),
    \vspace{-0.4mm}
\end{equation}
where $\mathcal{B}$ denotes a mapping function that removes all blank symbols from the alignment sequence $\pi$ such that $\mathcal{B}(\pi) = Y$.
The ST loss function $\mathcal{L}_\text{st}$ is obtained by replacing $Y$ with $Z$ in Eq.~\ref{eq:rnnt}. The final multi objective loss for training JSTAR is
\begin{equation}
\label{eq:jstar}
    \vspace{-0.8mm}
    \mathcal{L} = \mathcal{L}_\text{st} + \lambda \mathcal{L}_\text{asr}.
    \vspace{-0.4mm}
\end{equation}

As depicted in Figure~\ref{fig:jstar_arch}, the multi-channel audio input $\bm x_t$ is processed by a multi-channel front-end module similar to \cite{Lin2024AGADIR, lin2023directional}, which comprises multiple NLCMV-based beamformers, feature extraction, and a convolutional neural network (CNN) block.
In total, 13 NLCMV beamformers for 12 horizontal directions, in 30 degree steps around the glasses, and one mouth direction are applied.
Next, 80 dimensional Mel-spectral energy features are extracted for each of the 13 beamformer outputs.
Acoustic features are further processed by two consecutive 2D CNN layers each with 13 channels, a kernel size of $2 \times 5$, a GLU activation function and batch normalization.
Finally, a time reduction layer is applied which stacks 6 consecutive frames to form a 1326 dimensional feature vector that is projected to 320 dimension using a linear layer at a frame rate of 60~ms.

The output of the front-end module is then fed to the fast encoder, which is a stack of low-latency steaming conformer neural network layers \cite{Yangyang2022convemformer}.
The ASR predictor and joiner are used to generate ASR predictions.
The output of the fast encoder is input to the slow encoder, which is also based on a stack of streaming conformer layers with a larger latency configuration.
The final ST output is generated using the ST predictor and joiner, whose parameters are not shared with the ASR predictor and joiner.

\subsection{Multi-talker Training}
\label{ssec:model_arch}

JSTAR is applied for simultaneously recognizing and translating speech of a bilingual conversation with smart glasses in a streaming fashion.
In order to recognize multiple speakers with an RNN-T based system without explicit speaker separation, we are using SOT.
Let's consider the following case, where the wearer of the smart glasses (SELF) is speaking the English sentence ``Yesterday, I was talking to your sister Elizabeth'' and the conversational partner (OTHER) is responding in Spanish ``Genial, ¿qué dijo ella?''.
For SOT the temporal alignment of each word for both speakers must be estimated, e.g., via forced-alignment, such that the words of the target transcription can be sorted in chronological using the end time of each word.
To distinguish the speakers, a special label is added before each speaker segment in the SOT sequence.
For example, the SOT label sequence $Y$ for the ASR task would be
``$\langle$SELF$\rangle$ Yesterday, I was talking to your sister $\langle$OTHER$\rangle$ Genial, $\langle$SELF$\rangle$ Elizabeth. $\langle$OTHER$\rangle$ ¿qué dijo ella?'',
where the word ``Genial,'', which is spoken by OTHER, is temporally overlapping with SELF speech.

The corresponding SOT sequence $Z$ for the translation task is generated from $Y$ using an MT model, e.g.,
``$\langle$SELF$\rangle$ Ayer estaba hablando con tu hermana $\langle$OTHER$\rangle$ Great, $\langle$SELF$\rangle$ Elizabeth. $\langle$OTHER$\rangle$ what did she say?''.
In this case, text-to-text alignments are also required to obtain the correct word ordering for overlapping speech conditions as well as for conversational style data generation when utterances are segmented and mixed to generate more speaker turns.
For instance, such textual translation alignments can be generate using awesome-align \cite{awesomealign2021}.
In this work, we're leveraging the attention weights produced by an MT model instead.

\vspace{-0.5mm}
\section{Transducer-based Machine Translation}
\label{sec:MT}
\vspace{-0.5mm}

\begin{figure}[t]
  \centering
  \centerline{\includegraphics[width=0.8\linewidth]{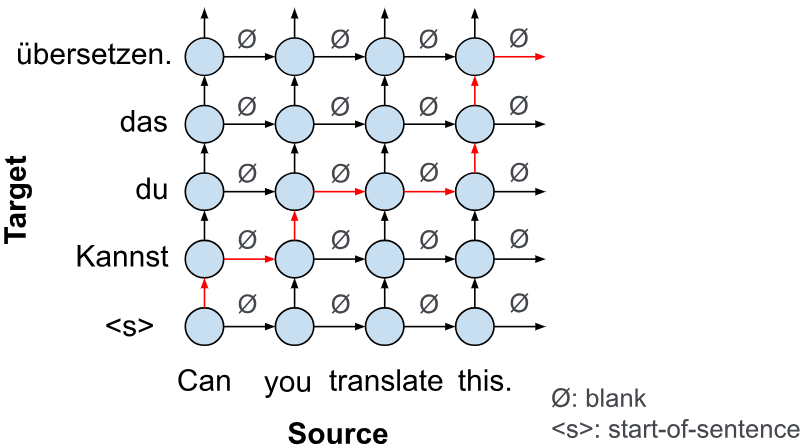}} 
  \vspace{-2mm}
  \caption{RNN-T lattice for an MT task example.}
  \vspace{-5mm}
\label{fig:rnnt_lattice}
\end{figure}

Typically, attention-based encoder-decoder (AED) architectures are applied for the MT task.
However, such architectures are difficult to apply for streaming translation, where the translated text is generated as soon as sufficient input words are available instead of waiting for a whole sentence. Efficient Monotonic Multi-head Attention (EMMA) \cite{seamless2023} is one solution to achieve this but is difficult to train for robust low latency setups.
For RNN-T, however, the continuous streaming case is much simpler and more natural achievable.
Typically, RNN-T is applied for ASR but it can be used for non-monotonic input-output alignments as well.
As an example, let’s assume we need to translate the English sentence “Can you translate this.” to German, which is “Kannst du das übersetzen.”. The word alignment between both sentences is not monotonic:

\begin{figure}[h]
  \centering
  \vspace{-4mm}
  \centerline{\includegraphics[width=0.46\linewidth]{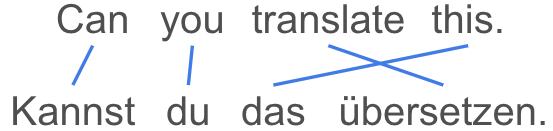}} 
  \vspace{-3mm}
\label{fig:word_ordering}
\end{figure}

A corresponding RNN-T lattice \cite{Graves12rnnt} is shown in Figure~\ref{fig:rnnt_lattice}, where an optimal alignment between the source and target sentence is shown in red.
Thus, the RNN-T loss is well suited for this problem, as long as the encoder in combination with the predictor neural network of the RNN-T model is capable of delaying an output such that the model can learn to emit words in the correct order, which is often the case.

\section{Experiments}
\label{sec:experiments}
\vspace{-1mm}

\begin{table}[t]
    \caption{BLEU [\%] scores of an AED and transducer-based MT model using realistic bilingual conversational test data for Spanish (ES) and English (EN).}
    \vspace{-2mm}
  \label{tab:results_mt}
  \centering
  \setlength{\tabcolsep}{2.0mm}
  \renewcommand\arraystretch{1.06}
  \resizebox{.49\linewidth}{!}{
  \begin{tabular}{l c c}
    \Xhline{3\arrayrulewidth}
    Model & ES$\rightarrow$EN & EN$\rightarrow$ES \\
    \hline
    AED & 32.3 & 29.9 \\
    RNN-T & 30.1 & 29.8 \\
    \Xhline{3\arrayrulewidth}
  \end{tabular}}
  \vspace{-4mm}
\end{table}

\subsection{Settings}
\label{ssec:settings}

In all experiments of this work, we are using 20 streaming conformer layer for the fast encoder and 10 for the slow encoder. The frame rate of the input of the fast encoder is $60$~ms as produced by the front-end.
The chunk size of the fast encoder layers amounts to 5 frames ($300$~ms), with 1 frame right-context ($60$~ms) and 10 frames left-context ($600$~ms)~\cite{Yangyang2022convemformer,Mahadeokar2022StreamingPT}. This setup results in an average theoretical latency of $300/2~\text{ms} + 60~\text{ms} = 210~\text{ms}$ for the fast encoder.
The slow encoder uses a chunk size of 10 frames instead ($600$~ ms), resulting in an average theoretical latency of $600/2~\text{ms} + 60~\text{ms} = 360~\text{ms}$.
The hidden dimensions, the number of attention heads, and the dimension of the feed-forward blocks of both the fast and slow encoder layers amount to 320, 4, and 2048, respectively.
The predictors for the ST and ASR tasks are each using a single 256-dimensional LSTM layer, dropout, layer norm, and a final linear projection layer with 768 output dimensions. Both joiner layers are using a ReLU activation function and a linear layer to project the input to a vocabulary size of 9001, including the blank token.
We set $\lambda = 0.5$ in Eq.~\ref{eq:jstar} for all experiments.
The total number of JSTAR model parameters amounts to 121M, including the beamformer and feature extraction.

For the \textit{transducer-based MT experiments}, we are using the same JSTAR model architecture and settings except for the front-end module, which is skipped, and instead of 20 fast encoder layers we are only using 10.
This is to use the slow encoder, predictor, and joiner of the MT model for the JSTAR parameter initialization experiments in Section~\ref{ssec:jstar_results}.
The input to the streaming MT model is character-based, which are converted into trainable embedding vectors before the fast encoder. Frame rate reduction is not applied to more closely resemble the higher frame rate of the ST task. Thus, the chunk size of the fast and slow encoder amounts to 5 and 10 characters, respectively, with one character right-context. The output vocabulary corresponds to 9000 word-pieces, similar to JSTAR.
The intermediate RNN-T loss after the fast encoder is simply producing word-pieces in the source language using $\lambda = 0.1$ during training.
The RNN-T loss of the slow encoder is to produce outputs in the target language.

\subsection{Datasets}
\label{ssec:adatasets}

The JSTAR model is trained using an in-house dataset of de-identified English and Spanish audio data with about 12k hours for each language, which is a collection of datasets from various sources. Parts of the datasets were collected from third-party vendors, where the content ranges from voice assistant commands to simulations of conversations between people. Another part corresponds to video data that is publicly shared by Facebook users. All data is single-channel audio, which is denoted as \textit{sup} for the supervised training data in the following reading.
In addition, we also apply an unsupervised training dataset for both English and Spanish when indicated, which includes about 500K hours of pseudo-labeled data using a teacher model for each language, denoted as \textit{unsup} in the experiments section.
In order to train JSTAR, we applied a teacher MT model to automatically translate the transcriptions of the sup and unsup datasets into the target language.

As real multi-channel training data of sufficient amounts is not available, all multi-channel training data is simulated for a smart glasses device with a 5-channel microphone array similar to \cite{Lin2024AGADIR} and the recent CHiME-8 MMCSG challenge \cite{zmolikova24_chime}.
We first recorded a total of about 500 room impulse responses (RIRs) for various directions and rooms around the glasses at distances of 1 and 2 meters.
We then simulate multi-channel training data by placing single-channel speech recordings in space for SELF (the wearer) and for OTHER (the conversational partner) and then simulate a conversation between SELF and OTHER with some overlap. The OTHER speaker is located at forward-facing angles of -60$^{\circ}$ to +60$^{\circ}$ and cross-talk is simulated from other directions. In~\cite{lin2023directional}, this configuration is labeled V4.
In addition, we simulated realistic speaker volumes based on sound pressure levels (SPLs).

For training of MT models, we use publicly available datasets, e.g., NLLB, OpenSubtitle, CCMatrix, Wikipedia \cite{schwenk2019ccmatrix, fan2021beyond, lison2016opensubtitles2016, wolk2014building} as well as internal datasets.

For evaluation, we choose internally collected bilingual real-life Spanish-English conversation (RealConv) and the publicly available FLEURS \cite{conneau2023fleurs} datasets.
FLEURS are typically single channel close-talk recordings. For this work, we apply the same multi-channel (MC), multi-talker data simulation to generate a conversation between SELF and OTHER, named MC-FLEURS in the further reading.

\subsection{Transducer-based MT Results}
\label{ssec:mt_results}

Table~\ref{tab:results_mt} shows a comparison of bilingual evaluation understudy (BLEU) scores between a strong traditional transformer-based MT model, which has about 110M parameters, and the proposed streaming transducer-based MT model, which has about 70M parameters.
The test data are in-house collected real bilingual conversations of Spanish (OHTER) and English (SELF) speakers using a smart-glasses device. The data is manually annotated and translated but the input to the MT models is the output of an ASR system.
The results show that for English to Spanish (ES$\rightarrow$EN) both models perform similarly while for Spanish to English (ES$\rightarrow$EN) the RNN-T results are 2.1 BLEU points lower. It is important to note that the transducer-based MT model uses a very restricted segment and right-context size of only $5+1$ characters.

\vspace{-0.5mm}
\subsection{JSTAR Results}
\label{ssec:jstar_results}
\vspace{-0.5mm}

\begin{table}[tb]
    \caption{JSTAR results for different model initialization (Init.) strategies and ASR positions using the MC-FLEURS test data (underwent data simulation for SELF and OTHER).}
    \vspace{-2mm}
  \label{tab:results_asr_init}
  \centering
  \setlength{\tabcolsep}{2.0mm}
  \renewcommand\arraystretch{1.06}
  \begin{tabular}{l  c c  c c c c}
    \Xhline{3\arrayrulewidth}
    & & & \multicolumn{2}{c}{$\downarrow$WER[\%]} & \multicolumn{2}{c}{$\uparrow$BLEU[\%]} \\
    \# & ASR & Init. & ES(other) & EN(self) & ES$\rightarrow$EN & EN$\rightarrow$ES \\
    \hline
    1 & - & -  & - & - & 19.65 & 17.76 \\ %
    2 & slow & -  & 7.9 & 14.6 & 19.26 & 17.64 \\ %
    3 & fast & -  & 8.8 & 16.0 & 19.69 & 18.13 \\ %
    4 & fast & fast  & 7.9 & 15.8 & 20.39 & 18.64 \\ %
    5 & fast & slow  & 8.6 & 16.1 & 20.25 & 18.34 \\ %
    6 & fast & fast+slow  & 8.4 & 15.5 & 20.52 & 18.72 \\ %
    \Xhline{3\arrayrulewidth}
  \end{tabular}
  \vspace{-4mm}
\end{table}

\begin{table}[tb]
    \caption{BLEU scores [\%] for JSTAR with sup+unsup training data using the RealConv and MC-FLEURS test datasets.}
    \vspace{-2mm}
  \label{tab:results_unsup}
  \centering
  \setlength{\tabcolsep}{2.0mm}
  \renewcommand\arraystretch{1.06}
  \begin{tabular}{l c  c c c c}
    \Xhline{3\arrayrulewidth}
    & & \multicolumn{2}{c}{RealConv} & \multicolumn{2}{c}{MC-FLEURS} \\
    \# & Init. & ES$\rightarrow$EN & EN$\rightarrow$ES & ES$\rightarrow$EN & EN$\rightarrow$ES \\
    \hline
    \multicolumn{2}{l}{cascaded baseline} & 45.4 & 43.9 & 21.6 & 19.3 \\
    \hline
    7 & - & 45.3 & 45.1 & 20.6 & 19.9 \\
    8 & fast & 45.5 & 45.5 & 20.6 & 19.9 \\
    9 & slow & 45.2 & 44.5 & 20.1 & 19.9 \\
    10 & fast+slow & 47.7 & 45.6 & 21.3 & 20.2 \\
    \Xhline{3\arrayrulewidth}
  \end{tabular}
  \vspace{-5mm}
\end{table}

The impact of ASR on ST performance is investigated by comparing system \#1 to \#3 of Table~\ref{tab:results_asr_init}, which are all using the same model architecture except for the position of the ASR joiner. System \#1 does not use any ASR loss ($\lambda=0$), \#2 applies the ASR joiner after the slow encoder, and \#3 after the fast encoder (as shown in Figure~\ref{fig:jstar_arch}).
System \#3 achieves the best BLEU score on the multi-channel (MC) FLEURS test data and similar speaker attributed (SA) word error rates (WER) \cite{chime8task3eval} using RealConv.
\#2 achieves the best ASR results due to using more layers and the higher latency setup from the slow encoder. However, sharing the slow encoder with the ST task impacts the ST performance, resulting in a regression in BLEU scores compared to system \#1 and \#3.

Next, the impact of initializing the fast and slow model parameters is investigated using a pre-trained multilingual ASR model for the fast model parts and using the transducer-based MT model of Section~\ref{ssec:mt_results} for the slow model parts, which includes the joiner and predictor for both ASR (fast) and ST (slow).
System \#4 in Table~\ref{tab:results_asr_init} shows that initialization of the fast model parameters improves both WER and BLEU scores of JSTAR, with average improvements of 0.55\% and 0.61\%, respectively.
The results of system \#5 show that initializing the slow model parts can improve BLEU scores while having a minimal impact on WERs. This is expected because the ASR-related model parameters remain uninitialized.
Using both fast+slow model initialization further helps to improve translation results, lifting BLEU scores from 19.69\% and 18.13\% (\#3) to 20.52\% and 18.72\%, corresponding to an average relative improvement of 3.7\%.

Table~\ref{tab:results_asr_init} results are obtained using supervised (sup) training data only, while results for the larger sup+unsup training data are presented in Table~\ref{tab:results_unsup}.
It is evident that the additional training data substantially improves ST results of JSTAR.
Moreover, fast (\#8) or slow (\#9) initialization alone no longer demonstrates effectiveness when compared to model \#7, which was trained from scratch.
However, initializing both fast and slow model parameters from pre-trained models still achieves substantial improvements. For example, BLEU points increase by 0.7 (ES$\rightarrow$EN) and 0.3 (EN$\rightarrow$ES) when using MC-FLEURS and by 2.4 (ES$\rightarrow$EN) and 0.5 (EN$\rightarrow$ES) for RealConv.

Table~\ref{tab:results_unsup} presents a comparison between JSTAR and a strong cascaded ASR+MT system, which utilizes a transformer-based non-streaming MT model.
On MC-FLEURS, JSTAR achieves a 0.9\% higher BLEU score for EN$\rightarrow$ES and a 0.3\% lower score for ES$\rightarrow$EN. Furthermore, on the real in-house test dataset (RealConv), JSTAR obtains an average of 2 extra BLEU points.

\subsection{Latency Evaluation}
\label{ssec:latency_eval}

\begin{table}[tb]
    \caption{P50 latency of the first and last finalized output token.}
  \label{tab:latency}
  \vspace{-2mm}
  \centering
  \setlength{\tabcolsep}{2.0mm}
  \renewcommand\arraystretch{1.06}
  \resizebox{.6\linewidth}{!}{
  \begin{tabular}{l c c}
    \Xhline{3\arrayrulewidth}
    Model & First Token & Last Token \\
    \hline
    Cascaded & 7.1 sec. & 2.6 sec. \\
    JSTAR & 3.3 sec. & 2.5 sec. \\
    \Xhline{3\arrayrulewidth}
  \end{tabular} }
  \vspace{-5mm}
\end{table}

We evaluate the ST latency of both the cascaded system and JSTART using RealConv test, which are recordings of real bilingual conversations, each several minutes long with multiple speaker turns. The average latency for emitting the first and last final token for both systems is presented in Table~\ref{tab:latency}.
Final tokens refer to those that cannot change anymore, typically due to \textit{N}-best reordering during beam search. The token finalization process and streaming decoding for using a non-streaming MT model in a cascaded system are described in \cite{rabatin2024}. Token finalization is important, e.g., to avoid flickering and for TTS playback. JSTAR exhibits a 3.8 sec. lower first token latency compared to the cascaded system. The latency for emitting and finalizing the last token is similar for both system, because ASR punctuation at sentence ends are used by the cascaded system for triggering the MT and finalization.

\vspace{-1mm}
\section{Conclusions}
\label{sec:conclusions}
\vspace{-0.5mm}

We demonstrated that the proposed fast+slow model architecture of JSTAR for simultaneous streaming ASR and ST can leverage both speech and text data for training and leads to optimal streaming translation results.
To the best of our knowledge, we also show for the first time that the transducer architecture is well suited for streaming MT, achieving competitive results to a transformer-based model. Additionally, we demonstrate the effectiveness of using the transducer-based MT model to initialize JSTAR parameters, resulting in improved ST results and outperforming a strong cascaded baseline system in both BLEU scores and token emission latency.

\bibliographystyle{IEEEbib}
\bibliography{strings,refs}

\end{document}